\documentclass[twocolumn,secnumarabic,amssymb, nobibnotes, aps, prd, superscriptaddress]{revtex4-1}

\PassOptionsToPackage{square,numbers}{natbib}

\usepackage{bm}
\usepackage{amsmath}
\usepackage{graphicx}
\usepackage[ruled]{algorithm2e}
\usepackage{amssymb}
\usepackage[makeroom]{cancel}

\DeclareMathOperator*{\argmax}{arg\,max}
\DeclareMathOperator*{\argmin}{arg\,min}
\DeclareMathOperator{\sign}{sign}

\begin{document}

\title{Optimally Influencing Complex Ising Systems}

\author{Christopher Lynn}
\affiliation{Department of Physics and Astronomy, University of Pennsylvania, Philadelphia, Pennsylvania 19104, USA}
\author{Daniel D. Lee}
\affiliation{Department of Electrical and Systems Engineering, University of Pennsylvania, Philadelphia, Pennsylvania 19104, USA}

\begin{abstract}
In the study of social networks, a fundamental problem is that of influence maximization (IM): How can we maximize the collective opinion of individuals in a network given constrained marketing resources? Traditionally, the IM problem has been studied in the context of contagion models, which treat opinions as irreversible viruses that propagate through the network. To study reverberant opinion dynamics, which yield complex macroscopic behavior, the IM problem has recently been proposed in the context of the Ising model of opinion dynamics, in which individual opinions are treated as spins in an Ising system. In this paper, we are among the first to explore the \textit{Ising influence maximization (IIM)} problem, which has a natural physical interpretation as the maximization of the magnetization given a budget of external magnetic field, and we are the first to consider the IIM problem in general Ising systems with negative couplings and negative external fields. For a general Ising system, we show analytically that the optimal external field (i.e., that which maximizes the magnetization) exhibits a phase shift from intuitively focusing on high-degree nodes at high temperatures to counterintuitively focusing on ``loosely-connected" nodes, which are weakly energetically bound to the ground state, at low temperatures. We also present a novel and efficient algorithm for solving IIM with provable performance guarantees for ferromagnetic systems in nonnegative external fields. We apply our algorithm on large random and real-world networks, verifying the existence of phase shifts in the optimal external fields and comparing the performance of our algorithm with the state-of-the-art mean-field-based algorithm.
\end{abstract}
\maketitle

\section{Introduction}

With the proliferation of online social networks over the last two decades, the problem of optimally influencing the opinions of individuals in social networks has garnered much attention \cite{Domingos-01, Richardson-01,Kempe-01}. The prevailing paradigm treats marketing as a viral process and is studied in the context of contagion models and irreversible processes \cite{Kempe-01,Mossel-01,Demaine-01,Goyal-01}. Under the viral paradigm, the influence maximization (IM) problem is, given a budget of seed infections, to choose the subset of individuals to infect such that the ensuing cascade is maximized at the completion of the viral process. While the viral paradigm accurately describes out-of-equilibrium phenomena, such as the introduction of new ideas or products to a system, this paradigm fails to capture reverberant opinion dynamics wherein repeated interactions between individuals in the network give rise to complex macroscopic opinion patterns, as, for example, is the case in the formation of political opinions \cite{Moussaid-01, Mas-01, Schelling-01, Isenberg-01,Galam-03}. In such cases, rather than maximizing the spread of a viral advertisement, the marketer is interested in optimally shifting the equilibrium opinions of individuals in the network.

\begin{figure*}[t]
\centering
\includegraphics[width=\textwidth]{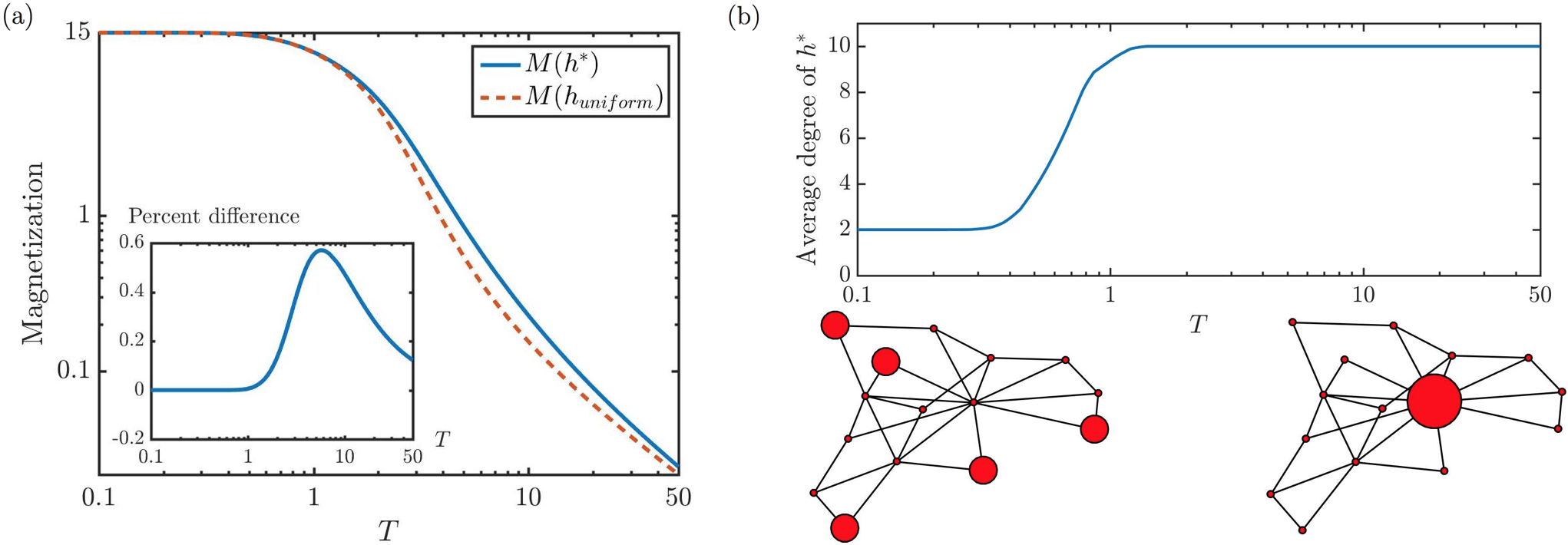}
\caption{\label{IntroExample} We consider a preferential attachment network consisting of $n=15$ nodes with ferromagnetic bonds of weight 1 and no initial external field. For a budget $H=1$, we can apply any additional external field $\bm{h}$ subject to the constraint $|\bm{h}|_1 \le H$. (a)  The magnetization of the system under the optimal external field $\bm{h}^*$ is compared with the magnetization under $\bm{h}_{uniform} = \frac{H}{n}(1,\hdots,1)^T$, revealing that the optimal external field yields up to 60\% larger magnetization. (b) The optimal external field undergoes a phase shift from spreading $H$ among the nodes of lowest degree at low temperatures to focusing on the node of highest degree at high temperatures.}
\end{figure*}

To study complex opinion formation, the influence maximization problem has recently been proposed in the context of the Ising model of opinion dynamics, in which individual opinions are treated as spins in an Ising system at dynamic equilibrium and marketing to the system is modeled as the addition of an external magnetic field \cite{Lynn-01,Liu-01}. The resulting problem, known as the \textit{Ising influence maximization (IIM)} problem, is equivalent to the maximization of the magnetization of an Ising system given a budget of external field. Figure \ref{IntroExample} gives an intuition for the IIM problem in a small preferential attachment network. In Figure \ref{IntroExample}(a) we find that, compared with the trivial strategy of spreading the external field uniformly among the nodes, focusing the external field on influential nodes can increase the magnetization of the system by up to 60\%. Remarkably, we find in Figure \ref{IntroExample}(b) that the structure of the optimal external field (i.e. that which maximizes the magnetization) undergoes a phase shift: at high temperatures, the magnetization is maximized by intuitively focusing on the single node of highest degree, while at low temperatures, the magnetization is maximized by counterintuitively spreading the external field among nodes of low degree.

In this paper, we explore the structure of optimal external fields in complex Ising systems, and we provide novel analytic and algorithmic tools for identifying influential nodes in large complex networks. One might suspect that the dramatic phase shift seen in Figure \ref{IntroExample}(b) is limited to a small class of Ising systems, and while previous work on the IIM problem has been limited to ferromagnetic systems in nonnegative external fields, we show analytically that such phase shifts are a persistent phenomenon found in all Ising systems, including those with negative couplings and negative external fields. In particular, for general Ising systems, we provide analytic solutions to IIM in the high-$T$ and low-$T$ limits, showing that the optimal external fields undergo a phase shift from focusing on high-degree nodes at high temperatures to focusing on ``loosely-connected" nodes, which are weakly energetically bound to the ground state, at low temperatures. 

We are also the first to study the Ising influence maximization problem directly (without mean-field approximations), and we provide a novel and efficient algorithm for solving IIM with provable performance guarantees for ferromagnetic systems in nonnegative external fields. We apply our algorithm on large random and real-world networks, verifying the existence of a phase shift in the optimal external fields and comparing the performance of our algorithm with the state-of-the-art mean-field-based algorithm in \cite{Lynn-01}.

The paper is organized as follows. In Section \ref{Model} we introduce the Ising model of opinion dynamics and formally state the IIM problem. In Sections \ref{HighT} and \ref{LowT} we derive analytic solutions to IIM in the the high-$T$ and low-$T$ limits, respectively. In Section \ref{Algorithms} we present an efficient algorithm for solving IIM, and we prove performance guarantees for the case of ferromagnetic systems in nonnegative external fields. In Section \ref{Experiments} we provide numerical experiments on large random and real-world networks, verifying the existence of phase shifts in the optimal external fields and comparing our algorithm with the state-of-the-art. We conclude with summarizing remarks in Section \ref{Conclusion}.

\subsection{Related work}

The recent development of the influence maximization problem can be traced to Domingos and Richardson \cite{Domingos-01,Richardson-01} and Kempe et al. \cite{Kempe-01}, the latter of which presented the problem in the context of the independent cascade model, marking the introduction of the viral paradigm that is still widely used today. Subsequent work has expanded the IM problem to include competing viruses \cite{Goyal-01}, fractional initial infections \cite{Demaine-01}, and continuous-time dynamics \cite{Gomez-Rodriguez-01,Du-01}, among many other interesting generalizations. However, all of this work is fundamentally based on contagion models, which do not allow for reverberant opinion dynamics and complex steady-states.

Our goal is to study the IM problem in contexts where individuals flip opinions as they interact, yielding complex collective behavior, and the Ising model is a natural starting point given its widespread success in describing complex systems. The Ising model has been applied to opinion formation dating to the work of Galam \cite{Galam-01,Galam-02}, and continues to be an active area of research \cite{Montanari-01,Castellano-01}. Ising-like models are also prevalent under names such as logistic response in game theory \cite{Blume-01,McKelvey-01} and the Sznajd model in sociophysics \cite{Sznajd-Weron-01}. Furthermore, complex Ising models have found widespread use in machine learning, and so the model we consider is formally equivalent to a pair-wise Markov random field or a Boltzmann machine \cite{Nishimori-01, Tanaka-01, Yedidia-01}. 

The Ising influence maximization problem represents a paradigm shift in the study of influence maximization and bridges a gap between the IM and Ising literatures. The IIM problem was formalized in \cite{Lynn-01}, where the authors limited their study to ferromagnetic systems in nonnegative external fields and only considered the IIM problem under the mean-field approximation. Our work represents two large steps forward: (i) we consider the IIM problem in general networks under general external fields, and (ii) we study the IIM problem directly via exact Ising calculations for small networks and Monte Carlo simulations for large networks. In order connect our results with previous work, we also develop a number of analytic results under the variational mean-field approximation, which has its roots in statistical mechanics and has long been used for inference in machine learning \cite{Jordan-01, Opper-01, Saul-01}. We note that Liu et al. also studied a similar problem in \cite{Liu-01}, but they neglected the effects of thermal noise, which we find to play a vital role.

\section{Ising influence maximization}
\label{Model}

In this section, we present the Ising model of opinion dynamics and describe how negative couplings and negative external fields account for two important features found in real-world social networks: (i) negatively correlated opinions and (ii) competitive marketing. We conclude by formally stating the Ising influence maximization problem in this generalized context.

\subsection{Generalized Ising model of opinion dynamics}

We consider an Ising system consisting of a set of individuals $N=\{1,\hdots ,n\}$, each of which is assigned an opinion $\sigma_i \in \{\pm 1\}$ that captures its current state. By analogy with the Ising model in statistical physics, we often refer to individuals as nodes and opinions as spins, and we refer to $\bm{\sigma}=(\sigma_i)$ as a spin configuration of the system. Individuals in the network interact via a symmetric coupling matrix $J$, where $J_{ij}=J_{ji}\in \mathbb{R}$, and we assume the diagonal entries in $J$ are zero such that self-interactions are forbidden. Each node also interacts with forces external to the network via an external field $\bm{h} = (h_i)$, where $h_i\in \mathbb{R}$ is the external field on node $i$. For example, if the spins represent the political opinions of individuals in a social network, then $J_{ij}$ represents the the amount of influence that individuals $i$ and $j$ hold over each other's opinions and $h_i$ represents the political bias of node $i$ due to external forces such as campaign advertisements and news articles.

We note that $\bm{h}$ and $J$ are allowed both positive and negative entries of arbitrary weight, whereas previous models have only considered $\bm{h}\ge 0$ and $J\ge0$. A negative external field $h_i <0$ represents a competitive external influence on node $i$. Competing sources of external influence are natural in most marketing contexts and have been studied extensively in the influence maximization literature \cite{Goyal-01,Bharathi-01}. A negative coupling $J_{ij}<0$ signifies that the spins $i$ and $j$ are anticorrelated, such that if individual $i$ has a positive opinion, then individual $j$ will be influenced to have a negative opinion, and vice versa. Anticorrelated opinions have been studied in the psychology literature \cite{Nowak-01}, but, to the authors' knowledge, have not yet been considered in models of influence maximization, even in the viral paradigm. Finally, we note that the introduction of negative couplings makes our model similar to a spin-glass, which is known to show much more complex behavior than a standard Ising ferromagnet \cite{Nishimori-02}.

The steady-state of an Ising system is described by the Boltzmann distribution
\begin{equation}
P(\bm{\sigma})=\frac{1}{Z}\exp\left(-\frac{1}{T}E(\bm{\sigma})\right),
\end{equation}
where $T$ is the temperature of the system, $E(\bm{\sigma}) = -\frac{1}{2}\sum_{ij}J_{ij}\sigma_i\sigma_j - \sum_i h_i\sigma_i$ is the energy of configuration $\bm{\sigma}$, and $Z = \sum_{\{\bm{\sigma}\}} \exp\left(-\frac{1}{T}E(\bm{\sigma})\right)$ is the partition function. Unless otherwise specified, sums over $i$ and $j$ run over the set of nodes $N$, and sums over $\{\bm{\sigma}\}$ are assumed over the set of spin configurations $\Omega=\{\pm 1\}^n$. The expectation of any function $f(\bm{\sigma})$ over (1) is denoted by $\left<f(\bm{\sigma})\right> = \sum_{\{\bm{\sigma}\}}f(\bm{\sigma}) P(\bm{\sigma})$. In particular, the total expected spin $M = \sum_i \left<\sigma_i\right>$ is referred to as the \textit{magnetization} of the system, and we will often consider the magnetization as a function of the external field, denoted by $M(\bm{h})$. 

Another important concept is the \textit{susceptibility}, represented by the matrix
\begin{equation}
\chi_{ij} = \frac{\partial \left<\sigma_i\right>}{\partial h_j} = \frac{1}{T}\left(\left< \sigma_i\sigma_j\right> - \left< \sigma_i\right>\left<\sigma_j\right>\right),
\end{equation}
where $\chi_{ij}$ quantifies the response of node $i$ to a change in the external field on node $j$. Identifying the susceptibility matrix with the spin covariance matrix, $\chi_{ij} = \frac{1}{T}\text{Covar}(\sigma_i,\sigma_j)$, it follows that $\chi$ is symmetric. Furthermore, to illustrate the importance of $\chi$, we note that, given vanishingly small external field resources $\delta H$, the magnetization is maximized by focusing $\delta H$ on the nodes corresponding to the largest elements of the gradient $\frac{\partial M}{\partial h_i} = \sum_j \chi_{ji}$. This intuition will prove useful in Section \ref{HighT}. \\

\noindent \textbf{Notation.} Unless otherwise specified, capital letters represent matrices and bold symbols represent column vectors with the appropriate number of components, while non-bold symbols with subscripts represent individual components. We often abuse notation and write relations such as $\bm{m}\ge 0$ to mean $m_i\ge 0$ for all components $i$. Along the same lines, any descriptor such as ``nonnegative," when applied to a vector or matrix, is assumed to be applied component-wise.

\subsection{Problem formulation}

We study the problem of maximizing the total magnetization of an Ising system with respect to the external field. For concreteness, we assume that an external agent can impose an external field $\bm{h}$ on the network subject to the constraint $|\bm{h}|_1 \le H$, where $|\bm{h}|_1=\sum_i |h_i|$ is the $\ell_1$ norm and $H>0$ is the \textit{external field budget}, and we denote the set of feasible external fields by $\mathcal{F}_H = \left\{ \bm{h} \in \mathbb{R}^n\, :\, |\bm{h}|_1 \le H\right\}$. In general, we also assume that the system experiences an \textit{initial external field} $\bm{h}^{(0)}$, which cannot be controlled by the external agent. We are now prepared to state the Ising influence maximization problem. \\

\noindent \textbf{Problem 1} (\textit{Ising influence maximization})\textbf{.} Given a system described by $J$, $\bm{h}^0$, and $T$, and an external field budget $H$, find an external field
\begin{equation}
\bm{h}^* = \argmax_{\bm{h}\in \mathcal{F}_H} M(\bm{h}^0+\bm{h}).
\end{equation}
We refer to any such external field as optimal. We also often refer to any nodes included in an optimal external field as optimal. \\

\noindent \textit{Remark.} Before proceeding, we point out that for finite systems, the objective $M(\bm{h}^0+\bm{h})$ is uniquely-defined for all $\bm{h}$ and hence the IIM problem is well-posed.

\section{High-$T$: high-degree nodes are optimal}
\label{HighT}

In this section, we develop an analytic solution to IIM in the high-temperature limit. For a general Ising system described by $J$ and $\bm{h}^{(0)}$ and an external field budget $H$, we prove in the $T\rightarrow\infty$ limit that the optimal external fields focus on the nodes with the highest degree in the network. While focusing on the nodes of highest degree is intuitive in that we tend to think of high-degree nodes as influential in driving network dynamics, we note that the independence of this result with respect to $\bm{h}^{(0)}$ is quite remarkable. We conclude by drawing comparisons with the external fields that maximize the mean-field magnetization, which have been the focus of previous work \cite{Lynn-01}.

\subsection{High-$T$ analytic solution to IIM}

In this subsection, we show that in the high-temperature limit, for a general Ising system and external field budget $H$, the magnetization is maximized by focusing $H$ on the nodes of highest degree. In what follows we consider a general Ising system defined by $J$, $\bm{h}$, and $T$. Instead of studying the magnetization directly, we find it most intuitive to consider the susceptibility $\chi$. In the high-$T$ limit, we can expand $\frac{1}{\beta}\chi_{ij}=\left<\sigma_i\sigma_j\right>-\left<\sigma_i\right>\left<\sigma_j\right>$ in powers of $\beta\equiv\frac{1}{T}$,
\begin{equation}
\label{TaylorExpansion}
\frac{1}{\beta}\chi_{ij} = \left(\frac{1}{\beta} \chi_{ij}\right)_{\beta=0} + \beta\frac{\partial}{\partial\beta}\left(\frac{1}{\beta} \chi_{ij}\right)_{\beta=0} + \hdots.
\end{equation}
Since the Boltzmann distribution is uniform at $\beta=0$, we have $\left<f(\bm{\sigma})\right>_{\beta=0} = \frac{1}{2^n}\sum_{\{\bm{\sigma}\}} f(\bm{\sigma})$ for any function of the spins $f(\bm{\sigma})$. Thus, when evaluated at $\beta=0$, any terms in $f(\bm{\sigma})$ that include an odd number of spins will vanish in expectation. In particular, $\left<\sigma_i\right>_{\beta=0}=0$ for all $i\in N$, and hence the zeroth-order term in Eq. (\ref{TaylorExpansion}) takes the form 
\begin{equation}
\label{X_beta}
\left(\frac{1}{\beta} \chi_{ij}\right)_{\beta=0} = \left<\sigma_i\sigma_j\right>_{\beta=0}=\frac{1}{2^n}\sum_{\{\bm{\sigma}\}}\sigma_i \sigma_j = \delta_{ij}.
\end{equation}

In Appendix A, we calculate the first- and second-order terms in Eq. (\ref{TaylorExpansion}), yielding
\begin{align}
\label{Xexpansion}
\chi_{ij} = &\,\,\beta\delta_{ij} \nonumber \\
&+ \beta^2 J_{ij} \nonumber \\
&+ \beta^3\left(\sum_k J_{ik}J_{kj} - \delta_{ij}\sum_k J_{jk}^2 - \delta_{ij}h_i^2\right) \nonumber \\
&+ \hdots.
\end{align}
Using Eq. (\ref{Xexpansion}), we can relate changes in the external field $\bm{h}$ to changes in the total magnetization $M$:
\begin{equation}
\label{Mexpansion}
\frac{\partial M}{\partial h_i} \equiv \sum_j \chi_{ij} = \beta + \beta^2 d_i + \beta^3\left(\sum_{j\neq i} (J^2)_{ij} - h_i^2\right) + \hdots,
\end{equation} 
where $d_i = \sum_j J_{ij}$ is the \textit{degree} of node $i$ and $\sum_{j\neq i}(J^2)_{ij} = \sum_{jk} J_{ik}J_{kj} - \sum_j J_{ij}^2$ represents the total weight of paths of length 2 originating from node $i$ (not counting self-interactions). 

Inspecting Eq. (\ref{Mexpansion}), we find to first-order in $\beta$ that $\frac{\partial M}{\partial h_i} > 0$, which tells us that in the high-$T$ limit, the optimal external fields $\bm{h}^*$ have positive components summing to $H$, i.e., $\sum_i h_i^*=H$. This may seem obvious, but we note that at intermediate and low temperatures, applying negative external field to nodes that are anticorrelated with the total magnetization can be optimal. To understand the dependence of the optimal external fields on the topology of the network, we turn to the second-order term in Eq. (\ref{Mexpansion}), which is proportional to the degree of node $i$. Since $\bm{h}^* \ge 0$ for all high-$T$ optimal external fields, we conclude that in the high-$T$ limit, the magnetization is maximized by focusing $H$ on the nodes of highest degree in the network, independent of the initial external field. The result is summarized in Theorem 2. \\

\noindent \textbf{Theorem 2.} Let $J$ and $\bm{h}^{(0)}$ describe a general Ising system and consider a general external field budget $H$. For sufficiently high temperatures, all optimal external fields place positive weight on the nodes of highest degree summing to $H$; i.e. letting $N_{max}=\argmax_i d_i$ denote the set of nodes of maximum degree, if $\bm{h}^*\in \mathcal{F}_H$ maximizes $M(\bm{h}^{(0)}+\bm{h})$ in the high-$T$ limit, then $\sum_{i\in N_{max}} h_i^*=H$, and hence $h^*_i=0$ for all $i\not\in N_{max}$. \\

Focusing one's budget on the nodes of highest degree is intuitive since we tend to think of high-degree nodes as influential in driving network dynamics. However, the independence of this result with respect to $\bm{h}^{(0)}$ can be counterintuitive, especially if one considers the extreme case of a network in which the node of highest degree experiences a large initial external field. 

If there exist multiple nodes of highest degree, then we turn to the third-order term in Eq. (\ref{Mexpansion}) to break the degeneracy between possible optimal external fields. The third-order term in Eq. (\ref{Mexpansion}) has a quadratic dependence on $\bm{h}$, favoring nodes with initial external fields that are small in magnitude. However, if $\bm{h}^{(0)}$ is uniform, then the high-$T$ optimal external fields focus $H$ on the nodes of highest degree which reach the largest number of nodes via paths of length 2. In the next subsection, we show that the mean-field magnetization is also maximized by focusing $H$ on the nodes of highest degree, but that the third-order term in Eq. (\ref{Mexpansion}) is different under the mean-field approximation, leading to the possibility of sub-optimal external fields at high temperatures.

\subsection{High-$T$ analytic solution under the mean-field approximation}

We now compare the optimal external fields to the mean-field optimal external fields (those which maximize the mean-field magnetization) in the high-temperature limit. We prove for general Ising systems that the high-$T$ mean-field magnetization is maximized by focusing on the nodes of highest degree, formally verifying a conjecture made in \cite{Lynn-01}. Furthermore, we go on to show that the third-order term in the high-$T$ expansion of the mean-field susceptibility is different from that in Eq. (\ref{Mexpansion}), leading to the possibility of sub-optimal external fields under the mean-field approximation.

For an Ising system described by $J$, $\bm{h}$, and $T$, the mean-field magnetization is defined by the self-consistency equations \cite{Jordan-01, Opper-01, Yedidia-01}:
\begin{equation}
\label{SelfConsistency}
m_i = \tanh\left[\beta\left(\sum_j J_{ij}m_j + h_i\right)\right],
\end{equation}
for all $i\in N$. To calculate the high-$T$ magnetization, we expand Eq. (\ref{SelfConsistency}) with respect to $\beta$:
\begin{equation}
\label{MFexpand1}
m_i = \beta\left(\sum_j J_{ij}m_j + h_i\right) - \frac{\beta^3}{3}\left(\sum_j J_{ij}m_j + h_i\right)^3 + \hdots.
\end{equation}
Because the leading-order term in $\bm{m}$ is first-order in $\beta$, we can drop $\sum_j J_{ij}m_j$ in the cubic term in Eq. (\ref{MFexpand1}). Abusing notation to let $\bm{h}^3 \equiv (h_1^3,\hdots, h_n^3)^T$, we have
\begin{align}
\bm{m} &= \beta(J\bm{m} + \bm{h}) - \frac{\beta^3}{3}\bm{h}^3 + \hdots \nonumber \\
\Rightarrow \quad \bm{m} &= (I - \beta J)^{-1}\left(\beta\bm{h} - \frac{\beta^3}{3}\bm{h}^3\right) + \hdots,
\end{align}
where $I$ is the $n\times n$ identity matrix. Assuming that our solution to Eq. (\ref{SelfConsistency}) is stable, i.e., $\bm{m}$ describes a minimum of the mean-field Gibbs free energy, we have that $\rho(\beta J) < 1$, where $\rho(\cdot)$ denotes spectral radius (see \cite{Lynn-01}). This allows us to expand $(I-\beta J)^{-1}$ as a geometric series of matrices, yielding
\begin{equation}
\label{MFexpand2}
\bm{m} = \beta \bm{h} + \beta^2 J\bm{h} + \beta^3\left(J^2\bm{h} -\frac{1}{3}\bm{h}^3\right) + O(\beta^4).
\end{equation}
From Eq. (\ref{MFexpand2}), one can easily derive the mean-field susceptibility of the system, which takes the form
\begin{align}
\label{MFexpand3}
\frac{\partial M_{MF}}{\partial h_i} = \beta + \beta^2 d_i + \beta^3\left(\sum_j (J^2)_{ij} - h_i^2\right) + \hdots,
\end{align}

\begin{figure}
\centering
\includegraphics[width = 0.48\textwidth]{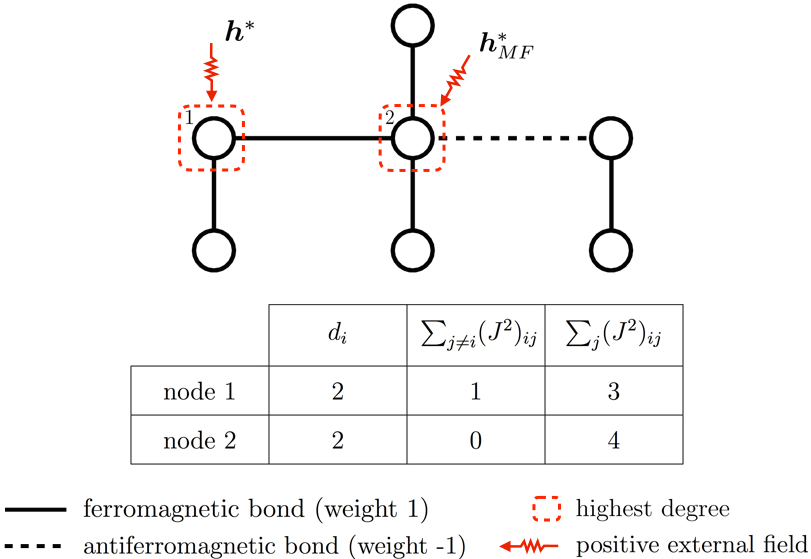}
\caption{\label{HighTExample} A small network in the high-temperature limit with $J_{ij}\in\{\pm 1\}$ corresponding to ferromagnetic and antiferromagnetic bonds, respectively. To second-order in $\beta$, nodes 1 and 2 are optimal, independent of $\bm{h}^{(0)}$ and $H$, since they both have maximum degree. However, the third-order term breaks the degeneracy: for $\bm{h}^{(0)}=0$ and $H<1$, the exact magnetization is maximized by focusing $H$ on node 1, while the mean-field magnetization is maximized by focusing $H$ on node 2.}
\end{figure}

Eq. (\ref{MFexpand3}) verifies, for a general Ising system and external field budget $H$, that under the mean-field approximation the high-$T$ optimal external fields are positive and correctly focus $H$ on the nodes of highest degree. However, the third-order term in Eq. (\ref{MFexpand3}) differs from the third-order term in Eq. (\ref{Mexpansion}) in that the sum over $(J^2)_{ij}$ runs over all $j$ instead of running over $j\neq i$. Thus, the mean-field approximation mistakenly takes into account self-interactions via paths of length 2, which can lead to sub-optimal external fields in the high-$T$ limit, as shown in Figure \ref{HighTExample}.

\section{Low-$T$: ``loosely-connected" nodes are optimal}
\label{LowT}

In this section, we study the structure of optimal external fields in the low-temperature limit. At low temperatures, the magnetization is governed by the spin configurations associated with low energies, i.e., the ground states, first-excited states, etc. However, for a large external field budget $H$, the feasible external fields $\bm{h}\in\mathcal{F}_H$ can alter the low-energy spectrum dramatically, making a general description of the low-temperature magnetization both cumbersome and unilluminating. To ease the discussion and foster intuition, we consider a natural special case: when the system admits a unique ground state and a unique first-excited state for all external fields $\bm{h}\in\mathcal{F}_H$. In this case, we show that the low-$T$ magnetization is maximized by focusing $H$ on the nodes with opposite parity between the ground state and first-excited state configurations, and we refer to such nodes as ``loosely-connected" since they are weakly energetically bound to the ground state. For an analytic discussion of low-temperature solutions to IIM for general Ising systems and general external field budgets we point the reader to Appendix B. In order to connect our results to previous work that has focused on maximizing the mean-field magnetization, we conclude this section by drawing comparisons with the low-$T$ mean-field optimal external fields.

\subsection{A low-$T$ analytic solution to IIM}

In this subsection, we show that in the low-temperature limit, for an Ising system that admits a unique ground state and a unique first-excited state for all $\bm{h}\in\mathcal{F}_H$, the magnetization is maximized by focusing $H$ on the nodes with opposite parity between the ground and first-excited states. We consider an Ising system described by $J$, $\bm{h}^{(0)}$, and $T$ and an external field budget $H$ such that the system admits unique ground and first-excited states $\bm{\sigma}^0$ and $\bm{\sigma}^1$, respectively, for all feasible external fields $\bm{h}\in\mathcal{F}_H$. In other words, we have that $\bm{\sigma}^0=\argmin_{\bm{\sigma}\in\{\pm1\}^n} E_{\bm{h}}(\bm{\sigma})$ and $\bm{\sigma}^1=\argmin_{\bm{\sigma}\in\{\pm1\}^n/\bm{\sigma}^0} E_{\bm{h}}(\bm{\sigma})$ for all $\bm{h}\in\mathcal{F}_H$, where $E_{\bm{h}}(\cdot)$ denotes the energy under the external field $\bm{h}^{(0)}+\bm{h}$. To give a sense of the systems for which this condition holds, we note that any system that initially (under $\bm{h}^{(0)}$) admits a unique ground state and first-excited state will continue to do so for all $\bm{h}\in\mathcal{F}_H$ as long as $H$ is small enough that the additional external field does not alter the low-energy spectrum, i.e., as long as $H<\frac{1}{2}\min\{\Delta E_{01}, \Delta E_{12}\}$ where $\Delta E_{01}$ and $\Delta E_{12}$ denote the initial energy gaps between the ground and first-excited states, and first-excited and second-excited states, respectively.

In the case described above, we find that the low-temperature magnetization is maximized by widening or narrowing the energy gap between $\bm{\sigma}^0$ and $\bm{\sigma}^1$. Under an external field $\bm{h}^{(0)}+\bm{h}$, for $\bm{h}\in\mathcal{F}_H$, the energy gap between the ground and first-excited states takes the form:
\begin{align}
\label{dE}
\Delta E(\bm{h}) &= E_{\bm{h}}(\bm{\sigma}^1) - E_{\bm{h}}(\bm{\sigma}^0) \nonumber \\
&= \Delta E_{01} +(\bm{\sigma}^0-\bm{\sigma}^1)^T\bm{h}.
\end{align}
For sufficiently low temperatures, we can write the total magnetization of the system in terms of $\Delta E(\bm{h})$:
\begin{align}
M(\bm{h}^{(0)}+\bm{h}) &\approx \frac{M_0 + M_1e^{-\frac{1}{T}\Delta E(\bm{h})}}{1+e^{-\frac{1}{T}\Delta E(\bm{h})}} \nonumber \\
&\approx M_0 + \left(M_1-M_0\right)e^{-\frac{1}{T}\Delta E(\bm{h})},
\end{align}
where $M_0=\sum_i\sigma_i^0$ and $M_1=\sum_i\sigma_i^1$ denote the magnetizations of the ground and first-excited states, respectively. Thus, the low-temperature optimal external fields, i.e., the external fields which maximize the low-temperature magnetization, take the form:
\begin{align}
\label{LowTh}
\bm{h}^* &= \argmax_{\bm{h}\in\mathcal{F}_H}\left(M_1-M_0\right)e^{-\frac{1}{T}\Delta E(\bm{h})} \nonumber \\
&\equiv \argmax_{\bm{h}\in\mathcal{F}_H} -\sign\left[M_1-M_0\right]\Delta E(\bm{h}) \nonumber \\
&\equiv \argmax_{\bm{h}\in\mathcal{F}_H} \sign\left[M_0-M_1\right](\bm{\sigma}^0-\bm{\sigma}^1)^T\bm{h}.
\end{align}

We first note that Eq. (\ref{LowTh}) is linear in $\bm{h}$ and is hence exactly solvable by efficient and scalable linear programing techniques. Thus, in the case of a unique ground state and a unique first-excited state, the efficiency of calculating the low-$T$ solutions to IIM is limited only by the computational complexity of identifying the ground and first-excited states. While calculating the ground state for a general Ising system is known to be NP-hard \cite{Barahona-01}, we note that there are efficient algorithms based on graph cuts for approximately minimizing the energy of general Ising systems within a known factor \cite{Boykov-01}. Furthermore, for ferromagnetic systems in arbitrary external fields, there are graph cut algorithms for provably and efficiently calculating the low-energy states \cite{Kolmogorov-01}. Since such graph-cut algorithms are primarily used in computer vision, Eq. (\ref{LowTh}) presents an interesting connection between IIM and computer vision problems in machine learning.

In addition to its interesting algorithmic properties, Eq. (\ref{LowTh}) also reveals a major insight into the structure of low-$T$ optimal external fields. Depending on the sign of $\left(M_0-M_1\right)$, the optimal external field $\bm{h}^*$ either maximizes the energy gap (in the case $M_0 >M_1$) or minimizes the energy gap (in the case $M_0 < M_1$). Furthermore, since the change in the energy gap equals $(\bm{\sigma}^0-\bm{\sigma}^1)^T\bm{h}$, it is clear that to maximize or minimize the energy gap, external field resources should be focused on nodes $i$ for which $\sigma_i^0\neq \sigma_i^1$. Thus, as long as $M_0\neq M_1$, the optimal external fields focus $H$ on the nodes with opposite parity between the ground and first-excited states. The result is summarized in Theorem 3. \\

\noindent \textbf{Theorem 3.} Consider an Ising system, described by $J$ and $\bm{h}^{(0)}$, for which there exist unique ground and first-excited states $\bm{\sigma}^0$ and $\bm{\sigma}^1$ for all feasible external fields $\bm{h}\in\mathcal{F}_H$, for some external field budget $H$. Letting $N_{LC} = \{i\in N\,:\, \sigma_i^0\neq \sigma_i^1\}$ denote the set of loosely-connected nodes with opposite parity between the ground and first-excited states, if $M_0 \neq M_1$, then for sufficiently low temperatures,  all optimal external fields $\bm{h}^*\in \mathcal{F}_H$ satisfy $\sum_{i\in N_{LC}} \left\vert h_i^*\right\vert=H$, and hence $h^*_i=0$ for all $i\not\in N_{LC}$. \\

Here we emphasize the structural dissimilarity between the high- and low-$T$ optimal external fields. For general Ising systems, at high temperatures, we found in Section \ref{HighT} that the optimal external fields focus on the nodes of highest degree in the network. We contrast this picture with the low-temperature limit, where, if the system admits unique ground and first-excited states for all $\bm{h}\in\mathcal{F}_H$, the optimal external fields focus on the loosely-connected nodes, which have opposite parity between the ground and first-excited states. In ferromagnetic systems, if the network is well-connected in the sense that the energy associated with flipping a set of nodes from the ground state tends to increase with the size of the set, then the first-excited state will differ from the ground state in only one or two entries, corresponding to nodes of low degree in the network. In Section \ref{Experiments}, we verify this intuition, showing on a number of random and real-world networks that the low-$T$ optimal external fields focus on nodes of low degree. Finally, we point the reader to Appendix B for an analytic description of the low-$T$ optimal external fields for general Ising systems and general external field budgets.

\subsection{A low-$T$ analytic solution under the mean-field approximation}

In order to connect the above results to previous work on the IIM problem, in this subsection we compare the optimal external fields to the mean-field optimal external fields in the low-temperature limit. As argued in the previous subsection, formulating a general description of the low-temperature mean-field magnetization is unilluminating. Thus, to ease the discussion, we limit our study to the mean-field analogue of the assumptions made in the previous subsection; namely, when the system admits a unique ground state and a unique node with the smallest effective field for all external fields $\bm{h}\in\mathcal{F}_H$. 

To clarify our assumptions, consider an Ising system described by $J$, $\bm{h}^{(0)}$, and $T$ and an external field budget $H$. Our first assumption is that there exists a unique ground state $\bm{\sigma}^0$ for all $\bm{h}\in \mathcal{F}_H$. For low temperatures, under an external field $\bm{h}^{(0)}+\bm{h}$, for $\bm{h}\in\mathcal{F}_H$, the mean-field magnetization of node $i$ takes the form:
\begin{align}
m_i &\approx \tanh\left[\frac{1}{T}\left(\sum_j J_{ij} \sigma_j^0 + h^{(0)}_i + h_i\right)\right] \nonumber \\
&\approx \sigma_i^0\left(1-2e^{-\frac{1}{T}2\sigma_i^0\left(\sum_j J_{ij} \sigma^0_j + h^{(0)}_i + h_i \right)}\right) \nonumber \\
&\equiv \sigma_i^0\left(1-2e^{-\frac{1}{T}2\tilde{h}_i}\right),
\end{align}
where $\tilde{h}_i = \sigma_i^0\left(\sum_j J_{ij} \sigma^0_j + h^{(0)}_i + h_i\right)$ is the \textit{effective field} felt by node $i$, and the second approximation follows by Taylor expansion. Since $\sigma_i^0 = \sign\left[\sum_j J_{ij} \sigma^0_j + h^{(0)}_i + h_i\right]$, we note that $\tilde{h}_i\ge 0$ for all $i\in N$. Furthermore, we note that the nodes $i$ for which $|m_i-\sigma_i^0|$ is greatest are the nodes with the smallest effective fields $\tilde{h}_i$. Thus, our second and final assumption is that there exists a unique node $i^* = \argmin_i \tilde{h}_i$ for all $\bm{h}\in\mathcal{F}_H$. As suggested by the notation, we will show that the low-$T$ mean-field magnetization is maximized by focusing on node $i^*$.

Under the above assumptions, and for an external field $\bm{h}^{(0)}+\bm{h}$, where $\bm{h}\in\mathcal{F}_H$, the low-temperature mean-field magnetization takes the form:
\begin{align}
M_{MF}(\bm{h}^{(0)}+\bm{h}) &\approx \sum_i \sigma_i^0\left(1-2e^{-\frac{1}{T}2\tilde{h}_i}\right) \nonumber \\
&\approx M_0 - 2\sigma_{i^*}^0e^{-\frac{1}{T}2\tilde{h}_{i^*}},
\end{align}
since all other terms will be exponentially suppressed in the $T\rightarrow 0$ limit. Thus, for an external field budget $H$, the low-temperature optimal external field under the mean-field approximation is given by:
\begin{align}
\bm{h}^*_{MF} &= \argmax_{\bm{h}\in\mathcal{F}_H} -2\sigma_{i^*}^0e^{-\frac{1}{T}2\tilde{h}_{i^*}} \nonumber \\
&\equiv \argmax_{\bm{h}\in\mathcal{F}_H} \sigma_{i^*}^0\tilde{h}_{i^*} \nonumber \\
&= \argmax_{\bm{h}\in\mathcal{F}_H} \sum_j J_{i^*j} \sigma^0_j + h^{(0)}_{i^*} + h_{i^*} \nonumber \\
&= H\bm{e}_{i^*},
\end{align}
where $\bm{e}_i\in\mathbb{R}^n$ denotes the vector of all zeros, but for a one in the $i^{\text{th}}$ conmponent. Thus, when there exists a unique ground state and a unique node $i^* = \argmin_i \tilde{h}_i$ for all $\bm{h}\in\mathcal{F}_H$, the low-temperature mean-field magnetization is maximized by focusing $H$ on node $i^*$.

\begin{figure}
\centering
\includegraphics[width = 0.5\textwidth]{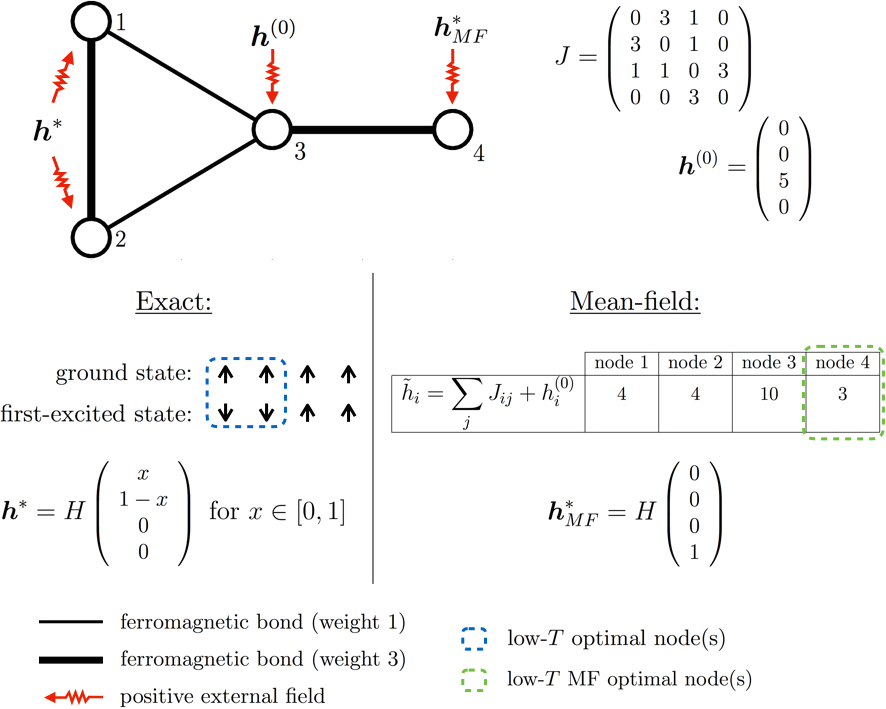}
\caption{\label{LowTMFComp} A small ferromagnetic network with bonds of weight 1 and 3 represented by thin and thick edges, respectively, under a positive initial external field $\bm{h}^{(0)} = (0,0,5,0)^T$. For $H<\frac{1}{2}$, the unique ground state is all-up $\bm{\sigma}^0 = (1,1,1,1)^T$ and the unique first-excited state is $\bm{\sigma}^1 = (-1,-1,1,1)^T$ for all $\bm{h}\in\mathcal{F}_H$. The low-$T$ optimal external fields apply $H$ to the loosely-connected nodes (nodes 1 and 2), while the mean-field optimal external field focuses $H$ on the node with the smallest effective field (node 4).}
\end{figure}

To illustrate the differences between $\bm{h}^*$ and $\bm{h}^*_{MF}$ at low temperatures, we consider the small ferromagnetic system shown in Figure \ref{LowTMFComp}, where thin edges represent bonds of weight 1, thick edges represent bonds of weight 3, and the system feels an initial external field $\bm{h}^{(0)}=(0,0,5,0)^T$. For $H<\frac{1}{2}$, the unique ground state is all-up $\bm{\sigma}^0 = (1,1,1,1)^T$, and the unique first-excited state is $\bm{\sigma}^1 = (-1,-1,1,1)^T$ for all $\bm{h}\in\mathcal{F}_H$. The low-$T$ optimal external fields $\bm{h}^*$ focus on the loosely-connected nodes with opposite parity between the ground and first-excited states (nodes 1 and 2). On the other hand, the mean-field low-$T$ optimal external field $\bm{h}^*_{MF}$ focuses on the node with the smallest effective field (node 4).  We also note for comparison that the high-$T$ optimal external field focuses on the node of highest degree (node 3), independent of $\bm{h}^{(0)}$ and $H$.

\section{A gradient ascent algorithm}
\label{Algorithms}

In addition to analytically exploring the structure optimal external fields in the high- and low-$T$ limits, we are also interested in developing efficient algorithmic tools for studying the structure of optimal external fields at all temperatures. In this section we present a novel algorithm that efficiently calculates a local maximum of $M(\bm{h}^{(0)} + \bm{h})$ in $\mathcal{F}_H$ for general Ising systems at arbitrary temperatures and for general external field budgets. For concreteness, by efficient we mean that, for any accuracy parameter $\epsilon>0$, our algorithm finds an $\epsilon$-approximation to a local maximum of $M(\bm{h}^{(0)} + \bm{h})$ in $O(1/\epsilon)$ iterations. In the special case of a ferromagnetic network ($J\ge 0$) in a nonnegative initial external field ($\bm{h}^{(0)}\ge 0$), we show that the magnetization is non-decreasing and concave in $\bm{h}$, and hence that our algorithm is guaranteed to converge to a global maximum of the magnetization, efficiently solving the IIM problem.

While previous work has focused on maximizing the mean-field magnetization \cite{Lynn-01}, our algorithm is the first to provide performance guarantees with respect to the exact magnetization. Our algorithm, summarized in Algorithm 1, is based on standard projected gradient ascent techniques. The algorithm is initialized at some feasible external field $\bm{h}^{(1)}\in\mathcal{F}_H$, and subsequently steps along the direction of the gradient of the magnetization, projected onto $\mathcal{F}_H$, until a local maximum is reached. In particular, at each iteration $t$, we calculate the gradient of the magnetization, $\frac{\partial M}{\partial h_i}= \sum_j \chi_{ij} = \frac{1}{T}\sum_j \left(\left<\sigma_i\sigma_j\right> - \left<\sigma_i\right>\left<\sigma_j\right>\right) $, under the external field $\bm{h}^{(0)}+\bm{h}^{(t)}$ and project the gradient onto the space of feasible external fields. We note that the projection operator, denoted $P_{\mathcal{F}_H}$, is well-defined since $\mathcal{F}_H$ is convex. Stepping along the direction of the projected gradient with step size $\alpha\in (0,\frac{1}{L})$, where $L$ is a Lipschitz constant of $M(\bm{h}^{(0)}+\bm{h})$ (which is well-defined since $M$ is smooth for finite systems), Algorithm 1 is known to converge to an $\epsilon$-approximation to a local maximum of $M(\bm{h}^{(0)}+\bm{h})$ in $O(1/\epsilon)$ iterations \cite{Teboulle-01}.
\begin{algorithm}[t]
\KwIn{An Ising system described by $J$, $\bm{h}^{(0)}$, and $T$; an external field budget $H$; and an accuracy parameter $\epsilon>0$}
\KwOut{An external field $\bm{h}$ that approximates a local maximum of $M(\bm{h}^{(0)}+\bm{h})$ in $\mathcal{F}_H$}
$t=1$; $\bm{h}^{(1)}\in\mathcal{F}_H$; $\alpha \in (0,\frac{1}{L})$ \;
\Repeat{$\left\vert M(\bm{h}^{(0)}+\bm{h}^{(t+1)}) - M(\bm{h}^{(0)}+\bm{h}^{(t)})\right\vert \le \epsilon$}{
$\; \frac{\partial M}{\partial h_i} = \sum_{j} \chi_{ij}$ for all $i\in N$ under $\bm{h}^{(0)} + \bm{h}^{(t)}$\;
$\;\bm{h}^{(t+1)} = P_{\mathcal{F}_H}\left[\bm{h}^{(t)} + \alpha \triangledown_{\bm{h}}M(\bm{h}^{(0)}+\bm{h}^{(t)})\right]$\;
$\;t$++\;
}
$\bm{h} = \bm{h}^{(t+1)}$;
\caption{Projected gradient ascent}
\label{Algorithm_1}
\end{algorithm}

We now show, in the special case of a ferromagnetic network ($J\ge 0$) in a nonnegative external field ($\bm{h}^{(0)}\ge0$), that the magnetization $M(\bm{h}^{(0)}+\bm{h})$ is non-decreasing and concave in $\bm{h}$, for $\bm{h}\ge 0$, and hence is efficiently maximized by Algorithm 1. The main result is summarized in Theorem 4. \\

\noindent \textbf{Theorem 4.} Consider a ferromagnetic network ($J\ge 0$) in a nonnegative initial external field ($\bm{h}^{(0)}\ge 0$) with arbitrary temperature $T>0$ and arbitrary external field budget $H>0$. Then, choosing $\bm{h}^{(1)}\ge 0$, Algorithm 1 converges to an $\epsilon$-approximation to IIM in $O(1/\epsilon)$ iterations for any accuracy parameter $\epsilon>0$. \\

To establish Theorem 4, we first note that the Fortuin-Kasteleyn-Ginibre inequality \cite{Fortuin-01} states that, for any ferromagnetic system, regardless of external field, $\chi_{ij} = \frac{\partial m_i}{\partial h_j} \ge 0$ for all $i,j\in N$, and hence $M(\bm{h}^{(0)}+\bm{h})$ is non-decreasing in $\bm{h}$. This implies two important facts: (i) there exists a global maximum in $\mathcal{F}_H$ for which $\bm{h}\ge 0$, and (ii) if Algorithm 1 is initialized such that $\bm{h}^{(1)}\ge 0$, then we will have $\bm{h}^{(t)}\ge 0$ for all subsequent $t$. Secondly, the Griffiths-Hurst-Sherman inequality \cite{Griffiths-02} guarantees that, for any ferromagnetic system in a nonnegative external field, $\frac{\partial^2 m_i}{\partial h_j\partial h_k} \le 0$ for all $i,j,k\in N$, and hence $M(\bm{h}^{(0)}+\bm{h})$ is concave in $\bm{h}$ for $\bm{h}\ge 0$. This implies that any local maximum of $M(\bm{h}^{(0)}+\bm{h})$ for $\bm{h}\ge 0$ is a global maximum for $\bm{h}\ge 0$ and hence, by fact (i) above, is a global maximum among all $\bm{h}\in \mathcal{F}_H$. Thus, in the case of a ferromagnetic system in a nonnegative initial external field, choosing $\bm{h}^{(1)}\ge 0$, Algorithm 1 will converge to a global maximum of $M(\bm{h}^{(0)}+\bm{h})$, efficiently solving the IIM problem. In the following section we apply Algorithm 1 on a number of random and real-world networks to study the structure of optimal external fields across a range of temperatures. \\

\noindent \textit{Remark.} While Algorithm 1 is efficient in that it converges to a local maximum in $O(1/\epsilon)$ iterations, the total run-time is limited by the calculation of $\chi$ and $M$ at each step. Since exact Ising calculations involve sums over $\{\pm 1\}^n$, which is exponential in the size of the system, an exact implementation of Algorithm 1 is limited to relatively small networks. To overcome this exponential dependence on system size, in the next section we use Monte Carlo simulations to approximate $\chi$ and $M$ at each iteration of Algorithm 1.

\section{Numerical experiments}
\label{Experiments}

In this section, we present numerical experiments to probe the structure of optimal external fields and to compare the performance of Algorithm 1 with the state-of-the-art mean-field-based algorithm in \cite{Lynn-01}. For a number of random and real-world networks, we verify our analytic results from Sections \ref{HighT} and \ref{LowT}, showing that the optimal external fields undergo phase shifts from focusing on high-degree nodes at high temperatures to focusing on low-degree nodes at low temperatures. 

Since an exact implementation of Algorithm 1 is limited to relatively small networks, to study large systems we employ an approximation to Algorithm 1 in which $\chi$ and $M$ are calculated via Monte Carlo simulations at each iteration. For simplicity, we refer to the exact implementation of Algorithm 1 as Algorithm 1(exact) and the Monte Carlo approximation as Algorithm 1(MC). On small random networks, we confirm that Algorithm 1(exact) outperforms both Algorithm 1(MC) as well as the mean-field-based algorithm in \cite{Lynn-01}, as expected. We also apply Algorithm 1(MC) on a number of large random and real-world networks, exhibiting its scalability and comparing the performance of Algorithm 1(MC) with the mean-field-based algorithm in \cite{Lynn-01}.

\subsection{Small networks}

We begin by studying the structure of optimal external fields, calculated exactly via Algorithm 1(exact), and comparing them with approximately optimal external fields, calculated via Monte Carlo simulations (Algorithm 1(MC)) and the mean-field approximation \cite{Lynn-01}. Since the algorithms we consider only have performance guarantees for ferromagnetic systems in nonnegative external fields, we limit our study to simple networks ($J_{ij}\in \{0,1\}$) with no initial external field ($\bm{h}^{(0)}=0$), and, for simplicity, we set $H=1$.

We first consider an Erd\"{o}s-R\'{e}nyi random network consisting of n=15 nodes, where each node is connected to each other node with probability $0.3$. 
\begin{figure*}
\centering
\includegraphics[width=\textwidth]{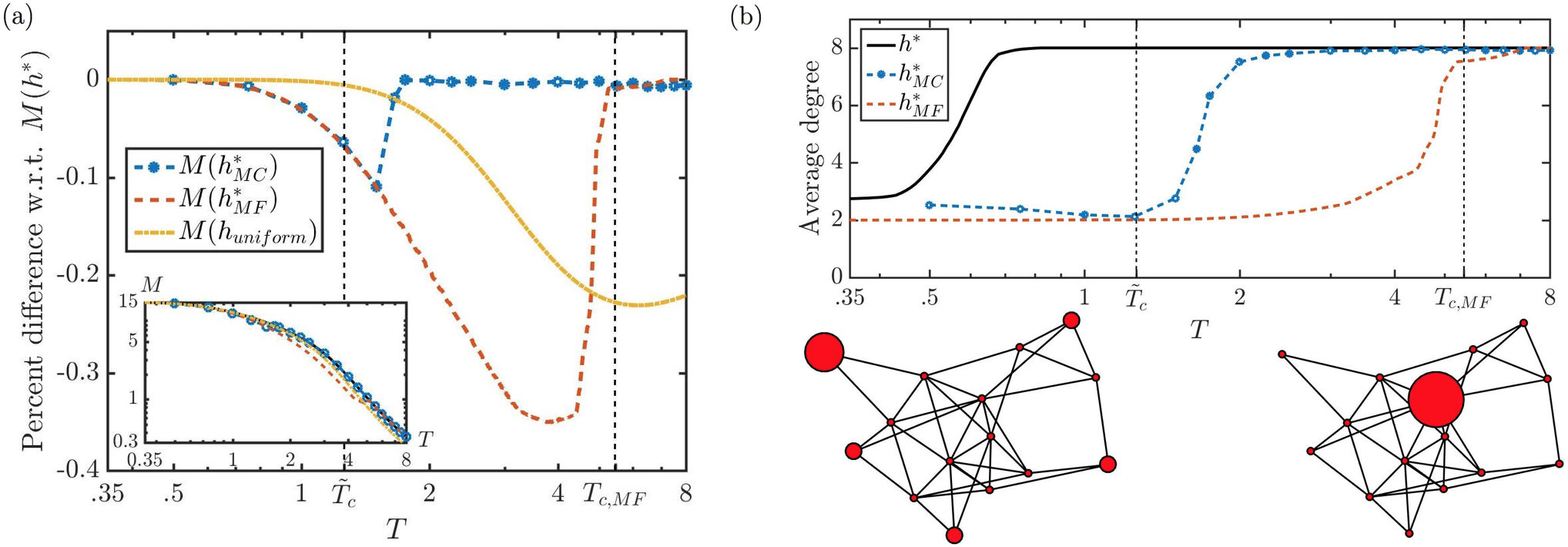}
\caption{\label{JRand} We consider an Erd\"{o}s-R\'{e}nyi random network composed of 15 nodes with edge probability 0.3 and couplings of weight 1, under initial external field $\bm{h}^{(0)}=0$. (a) For $H=1$, we compare $M(\bm{h}^*)$ with the magnetizations under $\bm{h}^*_{MC}$, $\bm{h}^*_{MF}$, and $\bm{h}_{uniform}$, confirming that $\bm{h}^*$ achieves the highest magnetization for all temperatures. (b) We verify that $\bm{h}^*$, $\bm{h}^*_{MC}$, and $\bm{h}^*_{MF}$ all undergo phase shifts from focusing on high- to low-degree nodes as the temperature decreases.}
\end{figure*}
The resulting system, shown in Figure \ref{JRand}, has maximum degree 8, minimum degree 2, and mean-field critical temperature $T_{c,MF} = \rho(J) = 5.45$ (see \cite{Lynn-01}), which sets the temperature scale of the system. For finite systems, the exact magnetization does not exhibit a phase transition; however, under a non-zero external field, any finite ferromagnetic system will exhibit a phase shift (a smooth transformation) marked by a peak in the susceptibility. We denote the temperature corresponding to maximum susceptibility by $\tilde{T}_c$, and for our Erod\"{o}s-R\'{e}nyi network with $H=1$ we have $\tilde{T}_c=1.26$.

In Figure \ref{JRand}(a), we compare the optimal magnetization, denoted $M(\bm{h}^*)$, with the magnetizations under external fields $\bm{h}^*_{MC}$, $\bm{h}^*_{MF}$, and $\bm{h}_{uniform}= \frac{H}{n}(1,\hdots,1)^T$, where $\bm{h}^*$ is calculated using Algorithm 1(exact), $\bm{h}^*_{MC}$ is calculated using Algorithm 1(MC) with Glauber dynamics consisting of 10,000 rounds of burn-in and 10,000 rounds of averaging at each iteration \cite{Levine-01}, and $\bm{h}^*_{MF}$ maximizes the mean-field magnetization \cite{Lynn-01}. We see clearly that $\bm{h}^*$ provides the largest magnetization, as expected, with $\bm{h}^*_{MC}$ admitting performance losses of up to 10\% and $\bm{h}^*_{MF}$ suffering performance losses of up to 35\% at intermediate temperatures. Furthermore, we find that Algorithm 1(MC) performs worst near $\tilde{T}_c$, likely due to critical slowing down of the Monte Carlo dynamics \cite{Newman-02}. We also note that the external fields are all near-optimal at high and low temperatures, since, for all $\bm{h}>0$, the magnetization approaches 0 as $T\rightarrow\infty$ and the magnetization approaches $n$ as $T\rightarrow 0$.

Figure \ref{JRand}(b) compares the phase shift in $\bm{h}^*$ with the phase shifts in $\bm{h}^*_{MC}$ and $\bm{h}^*_{MF}$ by calculating the average degrees of the external fields, defined by $\frac{\sum_i d_ih_i}{H}$, over a range of temperatures. We find that $\bm{h}^*$, $\bm{h}^*_{MC}$, and $\bm{h}^*_{MF}$ all exhibit phase shifts from focusing on high-degree nodes at high temperatures to focusing on low-degree nodes at low temperatures. However, each external field shifts at a different temperature, with $\bm{h}^*$ making the transition at the lowest temperature, followed by $\bm{h}^*_{MC}$ and then $\bm{h}^*_{MF}$ near $T_{c,MF}$.

\begin{figure*}
\centering
\includegraphics[width=\textwidth]{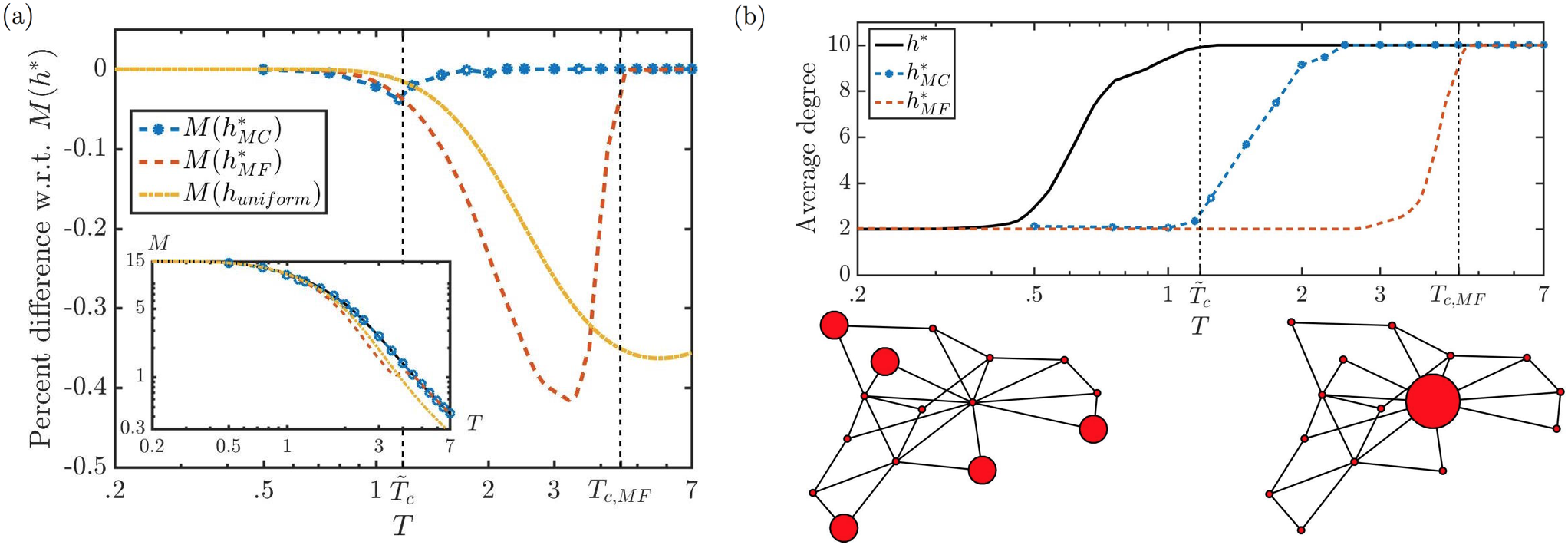}
\caption{\label{JPA} We consider a preferential attachment network composed of 15 nodes and couplings of weight 1, under initial external field $\bm{h}^{(0)}=0$. (a) For $H=1$, we compare $M(\bm{h}^*)$ with the magnetizations under $\bm{h}^*_{MC}$, $\bm{h}^*_{MF}$, and $\bm{h}_{uniform}$, verifying that $\bm{h}^*$ achieves the highest magnetization for all temperatures. (b) We verify that $\bm{h}^*$, $\bm{h}^*_{MC}$, and $\bm{h}^*_{MF}$ all undergo phase shifts from focusing on high- to low-degree nodes as the temperature decreases.}
\end{figure*}

We now consider a small preferential attachment network (also known as a Barab\'{a}si-Albert network) consisting of $n=15$ nodes, shown in Figure \ref{JPA}. 
Preferential attachment networks exhibit power-law degree distributions widely observed in social networks \cite{Albert-01}. Our system has a hub of maximum degree 10, five nodes of minimum degree 2, phase shift temperature $\tilde{T}_c=1.18$, and mean-field critical temperature $T_{c,MF}=4.51$. In Figure \ref{JPA}(a), we compare the magnetizations under external fields $\bm{h}^*$, $\bm{h}^*_{MC}$, $\bm{h}^*_{MF}$, and $\bm{h}_{uniform}$ over a range of temperatures. As in the Erod\"{o}s-R\'{e}nyi network, $\bm{h}^*$ achieves the highest magnetization for all temperatures, while $\bm{h}^*_{MC}$ performs second-best with performance losses limited to 4\% and localized near $\tilde{T}_c$, with $\bm{h}^*_{MF}$ suffering performance losses of up to 40\%. In Figure \ref{JPA}(b), we study the average degrees of $\bm{h}^*$, $\bm{h}^*_{MC}$, and $\bm{h}^*_{MF}$ over a range of temperatures, finding that all three exhibit phase shifts from focusing on the hub node at high temperatures to spreading $H$ among the minimum degree nodes at low temperatures. Furthermore, as in the Erod\"{o}s-R\'{e}nyi network, we find that $\bm{h}^*$ transitions at the lowest temperature, followed by $\bm{h}^*_{MC}$ and then $\bm{h}^*_{MF}$ near $T_{c,MF}$.

\subsection{Large networks}

We now shift our focus to large networks in order to exhibit the scalability of Algorithm 1(MC) and to study the structure of optimal external fields in large-scale systems. However, in scaling to large systems we lose two important tools: (i) we lose the use of Algorithm 1(exact) to exactly calculate optimal external fields, and (ii) we lose the ability to exactly calculate magnetizations and hence evaluate performance. Thus, in order to compare the performance of Algorithm 1(MC) with the mean-field-based algorithm in \cite{Lynn-01}, we calculate the magnetizations using 20 Glauber Monte Carlo runs with 5,000 rounds of burn-in and 10,000 rounds of averaging each. As in the previous subsection, we restrict our attention to simple networks ($J_{ij}\in\{\pm 1\}$) with no initial external field ($\bm{h}^{(0)}=0$).

\begin{figure*}
\centering
\includegraphics[width=\textwidth]{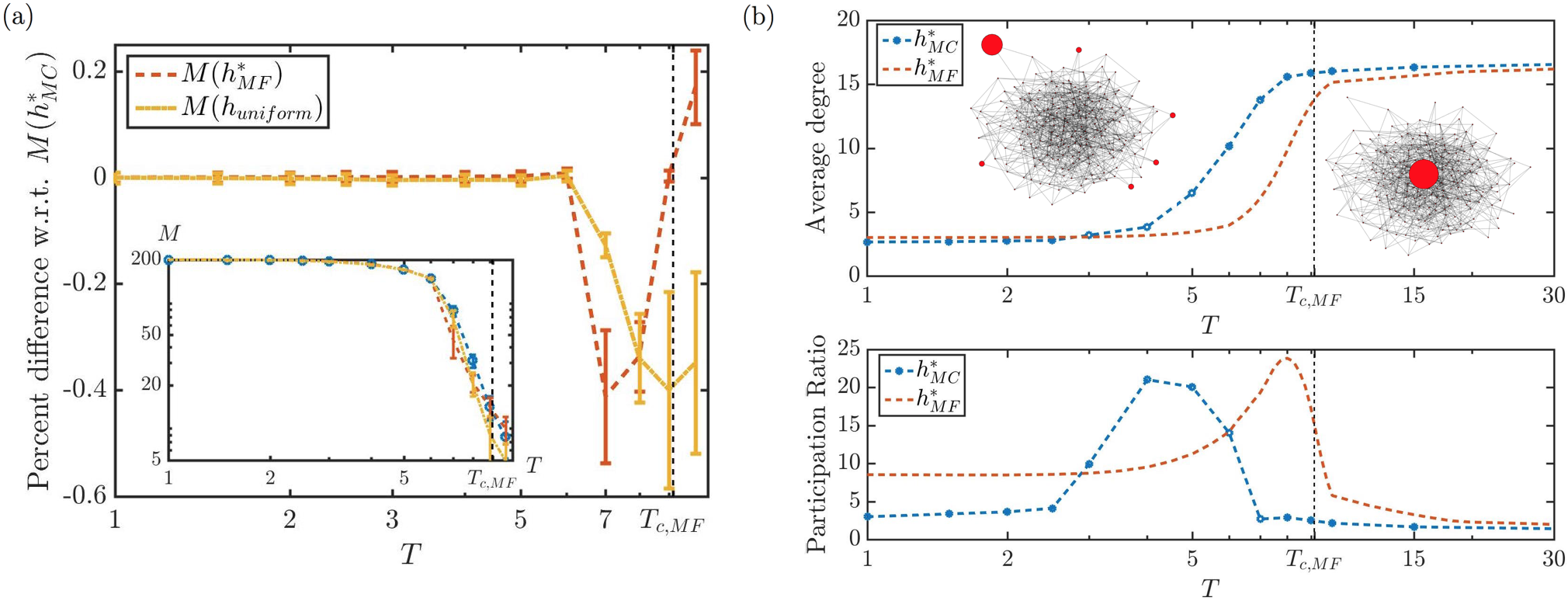}
\caption{\label{JRandBig} We consider an Erd\"{o}s-R\'{e}nyi random network composed of 200 nodes with edge probability 0.04 and couplings of weight 1, under initial external field $\bm{h}^{(0)}=0$. (a) For $H=10$, we compare the magnetizations under $\bm{h}^*_{MC}$, $\bm{h}^*_{MF}$, and $\bm{h}_{uniform}$, finding that $\bm{h}^*_{MC}$ achieves the highest magnetization for most temperatures. (b) We verify that $\bm{h}^*_{MC}$ and $\bm{h}^*_{MF}$ both undergo phase shifts from focusing on high- to low-degree nodes as the temperature decreases. Furthermore, we find for high and low temperatures that $\bm{h}^*_{MC}$ and $\bm{h}^*_{MF}$ only focus on a few nodes, while near the phase shifts they spread $H$ among many nodes.}
\end{figure*}

We first consider an Erod\"{o}s-R\'{e}nyi network, shown in Figure \ref{JRandBig}, consisting of $n=200$ nodes with edge probability 0.04. 
The resulting network has one node of maximum degree 17, one node of minimum degree 2, and mean-field critical temperature $T_{c,MF}=9.15$. In Figure \ref{JRandBig}(a), we compare of the magnetizations due to $\bm{h}^*_{MC}$, $\bm{h}^*_{MF}$, and $\bm{h}_{uniform}$ for an external field budget $H=10$, where the error bars represent one standard deviation from the mean for each batch of 20 Monte Carlo simulations. Across most temperatures we find that $\bm{h}^*_{MC}$ performs favorably compared with $\bm{h}^*_{MF}$, especially near $T_{c,MF}$, where $\bm{h}^*_{MF}$ admits performance losses of up to 40\%.

In Figure \ref{JRandBig}(b), we investigate the structure of $\bm{h}^*_{MC}$ and $\bm{h}^*_{MF}$ across a range of temperatures, specifically studying the average degree and the participation ratio, defined by $\frac{(\sum_i h_i)^2}{\sum_i h_i^2}$, which measures the effective number of nodes in the network over which $H$ is spread. As illustrated in the small networks, we find that both $\bm{h}^*_{MC}$ and $\bm{h}^*_{MF}$ exhibit phase shifts from focusing on high-degree to low-degree nodes as the temperature decreases, with $\bm{h}^*_{MF}$ transitioning near $T_{c,MF}$ and $\bm{h}^*_{MC}$ transitioning at a lower temperature. We also find for both high and low temperatures that $\bm{h}^*_{MC}$ and $\bm{h}^*_{MF}$ focus on only one or two nodes, while near the phase shifts they spread $H$ among many nodes. This indicates that the optimal external fields at intermediate temperatures involve many nodes not included in either the high-$T$ or low-$T$ optimal external fields.

\begin{figure*}
\centering
\includegraphics[width=\textwidth]{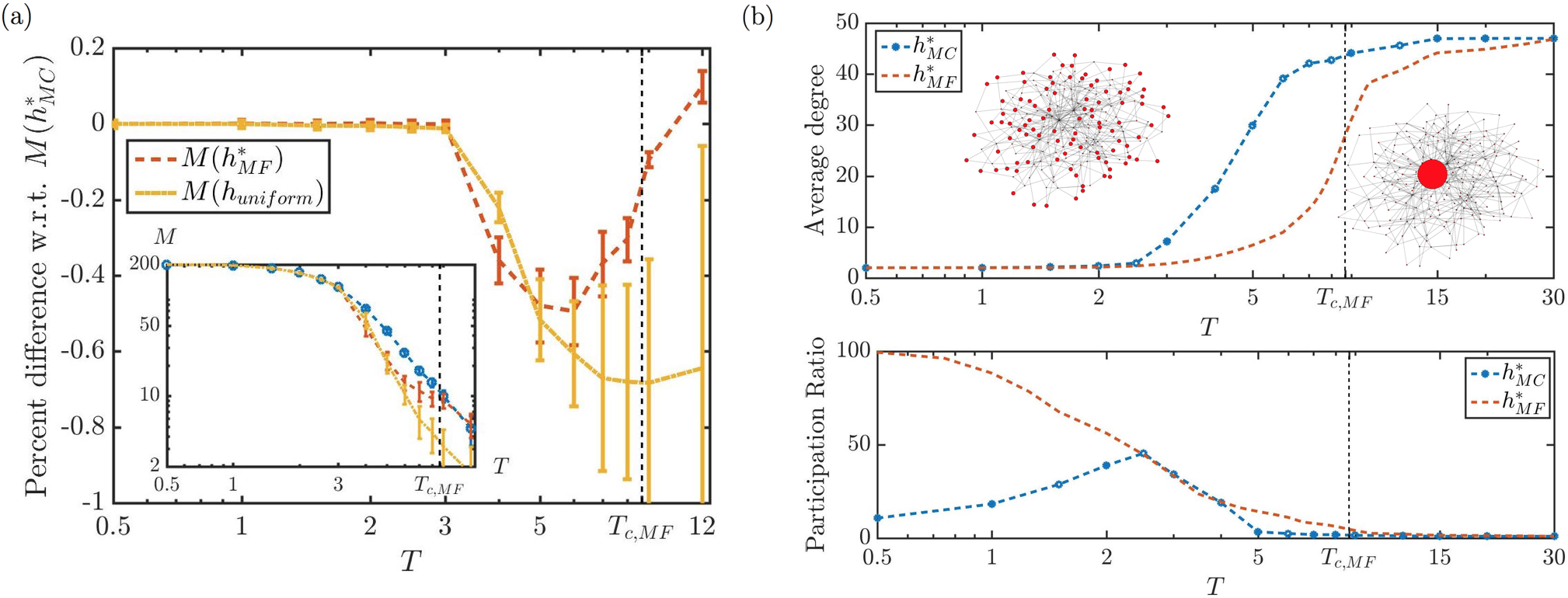}
\caption{\label{JPABig} We consider a preferential attachment network composed of 200 nodes and couplings of weight 1, under initial external field $\bm{h}^{(0)}=0$. (a) For $H=10$, we compare the magnetizations under $\bm{h}^*_{MC}$, $\bm{h}^*_{MF}$, and $\bm{h}_{uniform}$, finding that $\bm{h}^*_{MC}$ achieves the highest magnetization for most temperatures. (b) We verify that $\bm{h}^*_{MC}$ and $\bm{h}^*_{MF}$ both undergo phase shifts from focusing on high- to low-degree nodes as the temperature decreases. Furthermore, we find for high temperatures that both $\bm{h}^*_{MC}$ and $\bm{h}^*_{MF}$ focus on the node of highest degree, while at intermediate temperatures, $\bm{h}^*_{MC}$ spreads $H$ among a number of nodes. At low temperatures, $\bm{h}^*_{MF}$ spreads $H$ among the 99 nodes of lowest degree.}
\end{figure*}

Next, we consider a large preferential attachment network, shown in Figure \ref{JPABig}, consisting of $n=200$ nodes with one node of maximum degree 47, 99 nodes of minimum degree 2, and mean-field critical temperature $T_{c,MF}=8.67$. 
In Figure \ref{JPABig}(a), we compare the magnetization of the system under external fields $\bm{h}^*_{MC}$, $\bm{h}^*_{MF}$, and $\bm{h}_{uniform}$ for an external field budget $H=10$, finding that Algorithm 1(MC) compares favorably with the mean-field-based algorithm in \cite{Lynn-01} across most temperatures. In Figure \ref{JPABig}(b), we study the structure of $\bm{h}^*_{MC}$ and $\bm{h}^*_{MF}$ in the preferential attachment network, confirming that both external fields exhibit phase shifts from high- to low-degree nodes as the temperature decreases. Furthermore, looking at the participation ratio, we find that $\bm{h}^*_{MC}$ and $\bm{h}^*_{MF}$ both focus on the single node of highest degree at high temperatures, and, as in the  Erod\"{o}s-R\'{e}nyi network, $\bm{h}^*_{MC}$ spreads $H$ among many nodes near its phase shift. However, in contrast to the Erod\"{o}s-R\'{e}nyi network, $\bm{h}^*_{MF}$ continues to spread $H$ among an increasing number of nodes as the temperature decreases, eventually spreading $H$ among the 99 nodes of lowest degree.

\begin{figure*}
\centering
\includegraphics[width=\textwidth]{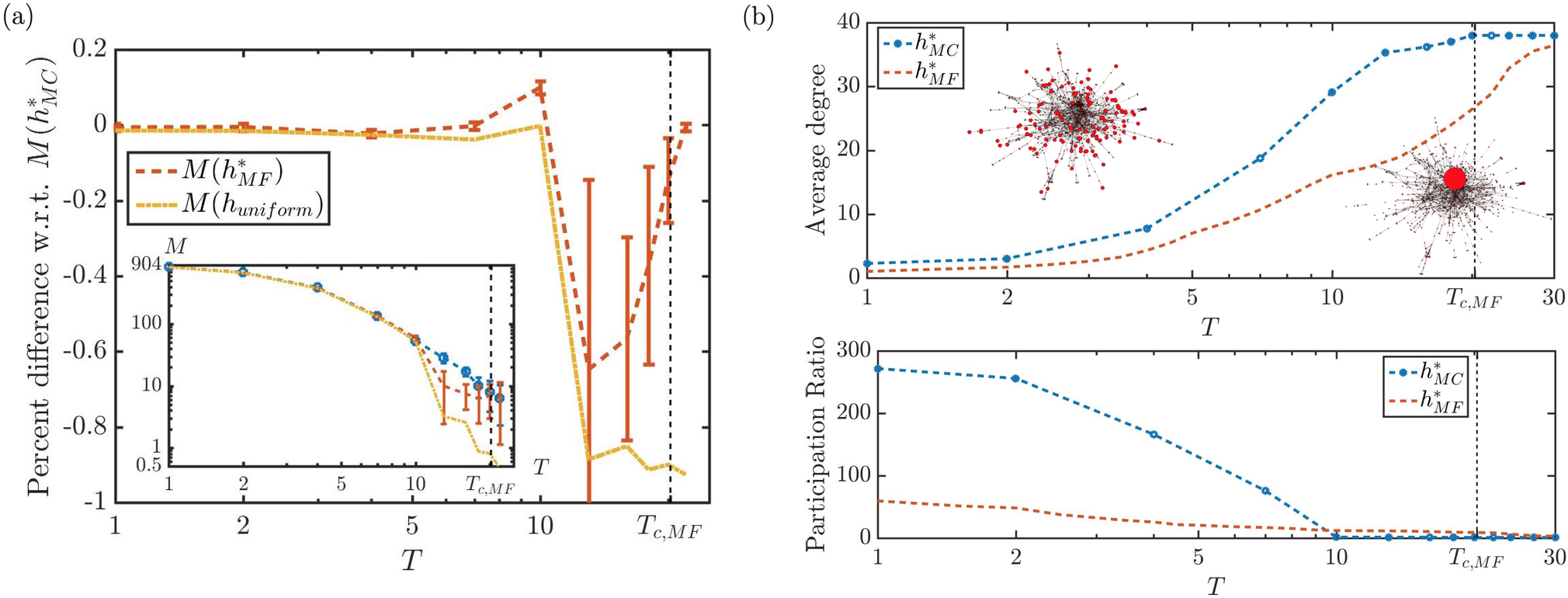}
\caption{\label{JReal} We consider a collaboration network consisting of 904 physicists where the edges represent co-authorships on papers posted to the arXiv. (a) For $H=20$, we compare the magnetizations under $\bm{h}^*_{MC}$, $\bm{h}^*_{MF}$, and $\bm{h}_{uniform}$, finding that $\bm{h}^*_{MC}$ performs well compared with $\bm{h}^*_{MF}$. (b) We confirm that $\bm{h}^*_{MC}$ and $\bm{h}^*_{MF}$ both undergo phase shifts from focusing on high-degree nodes to spreading $H$ among low-degree nodes as the temperature decreases.}
\end{figure*}

Finally, we consider a real-world collaboration network, shown in Figure \ref{JReal}, consisting of $n=904$ high-energy physicists and 2345 undirected, unweighted edges, where each edge represents the co-authorship of a paper posted to the arXiv \cite{SNAP-01}. 
We note that co-authorship networks are known to capture many of the key structural features of social networks \cite{Newman-01}, and our network has a maximum degree of 38, 127 nodes of minimum degree 1, and a mean-field critical temperature $T_{c,MF}=20.26$. In Figure \ref{JReal}(a), we compare the magnetization of the social network under external fields $\bm{h}^*_{MC}$, $\bm{h}^*_{MF}$, and $\bm{h}_{uniform}$ for an external field budget $H=20$, finding that Algorithm 1(MC) performs well compared with the mean-field based algorithm in \cite{Lynn-01}. We also note that the error bars for the percent difference of $M(\bm{h}_{uniform})$ with respect to $M(\bm{h}^*_{MC})$ have been intentionally omitted to improve the presentation. In Figure \ref{JReal}(b), we confirm that $\bm{h}^*_{MC}$ and $\bm{h}^*_{MF}$ undergo phase shifts from focusing on the node of highest degree at high temperatures to spreading $H$ among the nodes of lowest degree at low temperatures. Interestingly, at low temperatures we find that $\bm{h}_{MC}$ spreads $H$ among many more nodes than $\bm{h}_{MF}$.

\section{Conclusion}
\label{Conclusion}

In this paper we explore the Ising influence maximization problem, which has recently been proposed as a framework for studying the control and optimization of complex, reverberant opinion dynamics in large social networks. Traditionally, influence maximization in social networks has been studied in the context of contagion models and irreversible processes, so the IIM problem, which has a natural physical interpretation as the maximization of the magnetization given a budget of external magnetic field, marks an important departure from the existing paradigm and bridges a gap between the IM and Ising literatures. We are the first to consider the IIM problem in general Ising systems with negative couplings and negative external fields, accounting for anticorrelated opinions and competitive marketing, respectively, two important features found in real-world social networks.

In this paper we focus on describing the structure of optimal external fields (i.e. external fields which maximize the magnetization), and we provide novel analytic and algorithmic tools for identifying influential nodes in general complex Ising systems. Remarkably, we show analytically that the optimal external fields exhibit phase shifts from intuitively focusing on high-degree nodes at high temperatures to counterintuitively focusing on ``loosely-connected" nodes, which are weakly energetically bound to the ground state, at low temperatures. In order to connect our results with previous work that has focused on maximizing the mean-field magnetization, we also provide novel analytic descriptions of the mean-field optimal external fields, showing that they too exhibit a phase shift from focusing on high-degree nodes at high temperatures to focusing on the nodes with smallest effective fields at low temperatures.

We are also the first to study the Ising influence maximization problem directly (without mean-field approximations), and we provide a novel and efficient algorithm for finding local maxima of the magnetization in general Ising systems. Furthermore, for ferromagnetic systems in non-negative initial external fields, we prove that our algorithm converges to a global maximum, and hence efficiently solves the IIM problem. Finally, we apply our algorithm on large random and real-world networks, verifying the existence of phase shifts in the optimal external fields and comparing the performance of our algorithm with the state-of-the-art mean-field-based algorithm in \cite{Lynn-01}.

We believe that this paper opens the door for many exciting directions for future work. Many of the analytic and algorithmic tools developed to study Ising networks, ranging from renormalization group methods to advanced mean-field techniques, can now be applied in a wide array of contexts to study influence maximization in social networks. Furthermore, we believe that the IIM problem opens a new chapter for the Ising model by introducing a framework for exploring the control and optimization of complex Ising systems.

\section*{Acknowledgements}

We acknowledge support from the U.S. National Science Foundation, the Air Force Office of Scientific Research, and the Department of Transportation.

\section*{Appendix A: High-$T$ Susceptibility}

In what follows, we derive the first- and second-order terms in the high-temperature expansion of the susceptibility $\frac{1}{\beta}\chi_{ij}=\left<\sigma_i\sigma_j\right> - \left<\sigma_i\right>\left<\sigma_j\right>$ in Eq. (\ref{TaylorExpansion}). As noted in the main text, for any function of the spins $f(\bm{\sigma})$ we have
\begin{equation}
\left<f(\bm{\sigma})\right>_{\beta=0} = \frac{1}{2^n}\sum_{\{\bm{\sigma}\}}f(\bm{\sigma}),
\end{equation}
where the sum runs over all spin configurations. We also note that
\begin{align}
\frac{1}{2^n} \sum_{\{\bm{\sigma}\}} \sigma_i\sigma_j &= \delta_{ij},  \\
\frac{1}{2^n}\sum_{\{\bm{\sigma}\}} \sigma_i\sigma_j\sigma_k\sigma_{\ell} &= \delta_{ij}\delta_{k\ell} + \delta_{ik}\delta_{j\ell} + \delta_{i\ell}\delta_{jk} - 2\delta_{ijk\ell} \nonumber \\
&:= C_{ijk\ell},  \\
\frac{1}{2^n}\sum_{\{\bm{\sigma}\}} \sigma_i\sigma_j\sigma_k\sigma_{\ell}\sigma_m\sigma_p &= \delta_{ij}(\delta_{k\ell}\delta_{mp} + \delta_{km}\delta_{\ell p} + \delta_{kp}\delta_{\ell m}) \nonumber \\
&+\delta_{ik}(\delta_{j\ell}\delta_{mp} + \delta_{jm}\delta_{\ell p} + \delta_{jp}\delta_{\ell m}) \nonumber \\
&+\delta_{i\ell}(\delta_{jk}\delta_{mp} + \delta_{jm}\delta_{k p} + \delta_{jp}\delta_{k m}) \nonumber \\
&+\delta_{im}(\delta_{jk}\delta_{\ell p} + \delta_{j\ell}\delta_{k p} + \delta_{jp}\delta_{k \ell}) \nonumber \\
&+\delta_{ip}(\delta_{jk}\delta_{\ell m} + \delta_{j\ell}\delta_{k m} + \delta_{jm}\delta_{k \ell}) \nonumber \\
& - 2\left[ \delta_{ij}\delta_{k\ell mp} + \delta_{ik}\delta_{j\ell mp} + \delta_{i\ell}\delta_{jkmp}\right. \nonumber \\
& +\delta_{im}\delta_{jk\ell p} + \delta_{ip}\delta_{jk\ell m} + \delta_{jk}\delta_{i\ell mp}\nonumber \\
& +\delta_{j\ell}\delta_{ikmp} + \delta_{jm}\delta_{ik\ell p} + \delta_{jp}\delta_{ik\ell m}\nonumber \\
& +\delta_{k\ell}\delta_{ijmp} + \delta_{km}\delta_{ij\ell p} + \delta_{kp}\delta_{ij\ell m}\nonumber \\
& \left.+\delta_{\ell m}\delta_{ijkp} + \delta_{\ell p}\delta_{ijkm} + \delta_{mp}\delta_{ijk\ell}\right]\nonumber \\
&+16 \delta_{ijk\ell mp} \nonumber \\
&:= C_{ijk\ell mp},
\end{align}
while any expectation involving an odd number of spins vanishes. Finally, we note that for any function of the spins $f(\bm{\sigma})$, we have
\begin{equation}
\frac{\partial}{\partial \beta}\left<f(\bm{\sigma})\right> = -\left<f(\bm{\sigma})E(\bm{\sigma})\right> + \left<f(\bm{\sigma})\right> \left<E(\bm{\sigma})\right>,
\end{equation}
where $E(\bm{\sigma}) = -\frac{1}{2}\sum_{ij} J_{ij}\sigma_i\sigma_j - \sum_i h_i \sigma_i$ is the energy of configuration $\bm{\sigma}$.

We proceed with the the first-order term in Eq. (\ref{TaylorExpansion}). We recall that $\left<\sigma_i\right>_{\beta=0}=0$ for all $i\in N$ and we also have
\begin{align}
\left<E(\bm{\sigma})\right>_{\beta=0} &= -\frac{1}{2^n}\sum_{\{\bm{\sigma}\}}\left(\frac{1}{2}\sum_{ij}J_{ij}\sigma_i\sigma_j +\sum_i h_i \sigma_i\right) \nonumber \\
&= -\frac{1}{2}\sum_{ij}J_{ij}\delta_{ij} = 0,
\end{align}
since $J_{ii}=0$ for all $i\in N$ by definition. Thus, the first-order term in Eq. (\ref{TaylorExpansion}) takes the form
\begin{align}
\label{X_beta2}
\frac{\partial}{\partial \beta}\left(\frac{1}{\beta}\chi_{ij}\right)_{\beta=0} &= \left[-\left<\sigma_i\sigma_j E(\bm{\sigma})\right> + \cancel{\left< \sigma_i\sigma_j\right>\left<E(\bm{\sigma})\right>}\right. \nonumber \\
&\;\;\;\; -\cancel{\left<\sigma_i\right>\left(-\left<\sigma_j E(\bm{\sigma})\right> + \left<\sigma_j\right>\left<E(\bm{\sigma})\right>\right)} \nonumber \\
&\;\;\;\; \left.-\cancel{\left<\sigma_j\right>\left(-\left<\sigma_i E(\bm{\sigma})\right> + \left<\sigma_i\right>\left<E(\bm{\sigma})\right>\right)}\right]_{\beta=0} \nonumber \\
&= \frac{1}{2^n} \sum_{\{\bm{\sigma}\}}\sigma_i\sigma_j\left(\frac{1}{2}\sum_{k\ell}J_{k\ell}\sigma_k\sigma_{\ell} +\sum_k h_k \sigma_i\right) \nonumber \\
&= \frac{1}{2}\sum_{k\ell}J_{k\ell}\left( \delta_{ij}\delta_{k\ell} + \delta_{ik}\delta_{j\ell} + \delta_{i\ell}\delta_{jk} - 2\delta_{ijk\ell}\right) \nonumber \\
&= \frac{1}{2}\left(J_{ij}+J_{ji}\right) = J_{ij},
\end{align}
where the penultimate equality follows from $J$ being zero on the diagonal and the final equality follows since $J$ is symmetric.

We now derive the second-order term in Eq. (\ref{TaylorExpansion}). Including only the non-vanishing terms, we have
\begin{align}
\label{thirdOrder}
\frac{\partial^2}{\partial \beta^2}\left(\frac{1}{\beta}\chi_{ij}\right)_{\beta=0} &= \left[\left<\sigma_i\sigma_j E^2(\bm{\sigma})\right> - \left< \sigma_i\sigma_j\right>\left<E^2(\bm{\sigma})\right>\right. \nonumber \\
&\;\;\;\; \left.- 2\left<\sigma_iE(\bm{\sigma})\right>\left<\sigma_jE(\bm{\sigma})\right>\right]_{\beta=0}.
\end{align}
Evaluating the second term in Eq. (\ref{thirdOrder}), we recall $\left<\sigma_i\sigma_j\right>_{\beta=0}=\delta_{ij}$ and we have
\begin{align}
\left<E^2(\bm{\sigma})\right>_{\beta=0} &= \frac{1}{2^n}\sum_{\{\bm{\sigma}\}}\left(\frac{1}{2}\sum_{k\ell} J_{k\ell}\sigma_k\sigma_{\ell} + \sum_k h_k \sigma_k\right) \nonumber \\
&\qquad \qquad \cdot \left(\frac{1}{2}\sum_{mp} J_{mp}\sigma_m\sigma_p + \sum_m h_m \sigma_m\right) \nonumber \\
&=\frac{1}{4}\sum_{k\ell mp} J_{k\ell}J_{mp}C_{k\ell mp} + \sum_{km}h_kh_m\delta_{km} \nonumber \\
&=\frac{1}{2}\sum_{k\ell}J_{k\ell}^2 + \sum_k h_k^2.
\end{align}
Furthermore, the third term in Eq. (\ref{thirdOrder}) is given by
\begin{align}
\left<\sigma_iE(\bm{\sigma})\right>_{\beta=0} &= -\frac{1}{2^n}\sum_{\{\bm{\sigma}\}}\sigma_i\left(\frac{1}{2}\sum_{k\ell} J_{k\ell}\sigma_k\sigma_{\ell} + \sum_k h_k \sigma_k\right) \nonumber \\
&=-\sum_k h_k \delta_{ik} = -h_i.
\end{align}
Finally, we evaluate the first term in Eq. (\ref{thirdOrder}):
\begin{align}
\left<\sigma_i\sigma_j E^2(\bm{\sigma})\right>_{\beta=0} &=  \frac{1}{2^n}\sum_{\{\bm{\sigma}\}}\sigma_i\sigma_j\left(\frac{1}{2}\sum_{k\ell} J_{k\ell}\sigma_k\sigma_{\ell} + \sum_k h_k \sigma_k\right) \nonumber \\
&\qquad \qquad \cdot \left(\frac{1}{2}\sum_{mp} J_{mp}\sigma_m\sigma_p + \sum_m h_m \sigma_m\right) \nonumber \\
&=\frac{1}{4}\sum_{k\ell mp}J_{k\ell}J_{mp} C_{ijk\ell mp} + \sum_{k m} h_k h_m C_{ijkm} \nonumber \\
&= \frac{1}{2}\delta_{ij}\sum_{k\ell}J_{k\ell}^2 + 2\sum_k J_{ik}J_{kj} - 2\delta_{ij}\sum_k J_{jk}^2  \nonumber \\
& \qquad + \delta_{ij}\sum_k h_k^2 + 2h_ih_j - 2\delta_{ij}h_i^2.
\end{align}
Combining all of the terms in Eq. (\ref{thirdOrder}) and canceling appropriately, we are left with
\begin{align}
\label{X_beta3}
\frac{\partial^2}{\partial \beta^2}\left(\frac{1}{\beta}\chi_{ij}\right)_{\beta=0} &= 2\left(\sum_k J_{ik}J_{kj} - \delta_{ij}\sum_k J_{jk}^2 - \delta_{ij}h_i^2\right).
\end{align}
All together, Eqs. (\ref{X_beta}), (\ref{X_beta2}), and (\ref{X_beta3}) represent the zeroth-, first-, and second-order terms in the expansion of $\frac{1}{\beta}\chi_{ij}$ in Eq. (\ref{TaylorExpansion}), respectively. Multiplying by $\beta$ yields the high-temperature expansion of $\chi_{ij}$ in Eq. (\ref{Xexpansion}), as desired.

\section*{Appendix B: low-$T$ optimal external fields for general systems}

In this appendix, we provide an analytic description of the low-$T$ optimal external fields for a general Ising system, described by $J$ and $\bm{h}^{(0)}$, and a general external field budget $H$, generalizing the results of Section \ref{LowT}. For some external field $\bm{h}\in \mathcal{F}_H$, let $\Omega^0_{\bm{h}} = \argmin_{\{\bm{\sigma}\}} E_{\bm{h}}(\bm{\sigma})$ denote the set of ground state configurations under the external field $\bm{h}^{(0)} + \bm{h}$, where $E_{\bm{h}}(\cdot)$ denotes the energy under the external field $\bm{h}^{(0)} + \bm{h}$, and let $\Omega^0=\cup_{\bm{h}\in\mathcal{F}_H} \Omega^0_{\bm{h}}$ denote the set of all possible ground states that can be induced by some $\bm{h}\in\mathcal{F}_H$ for an external field budget $H$.

For sufficiently low temperatures, the ground state of the system dominates any expectation. Thus, any optimal external field will necessarily induce a ground state that has the largest magnetization among the configurations in $\Omega^0$. So, let ${\Omega^0}^*\subset \Omega^0$ denote the subset of possible ground states with the maximum magnetization and let $\mathcal{F}_H^* = \left\{\bm{h}\in \mathcal{F}_H\, :\, \Omega^0_{\bm{h}}\subset {\Omega^0}^*\right\}$ denote the set of feasible external fields that induce ground states of maximum magnetization.

For sufficiently low temperatures, any optimal external field $\bm{h}^*$ is located in $\mathcal{F}_H^*$. To differentiate between the external fields in $\mathcal{F}_H^*$, we consider the possible first-excited states, as in Section \ref{LowT}. Let $\Omega^1_{\bm{h}}$ denote the set of first excited states of the system under the external field $\bm{h}^{(0)}+\bm{h}$, and let $\Omega^1=\cup_{\bm{h}\in\mathcal{F}^*_H} \Omega^1_{\bm{h}}$ denote the set of possible first-excited states that can be induced by some $\bm{h}\in \mathcal{F}^*_H$. For $\bm{h}\in\mathcal{F}_H$, letting $\Delta E(\bm{h}) = \min_{\bm{\sigma}\in \Omega^1} \max_{\bm{\sigma}^0\in {\Omega^0}^*} \left(E_{\bm{h}}(\bm{\sigma}) - E_{\bm{h}}(\bm{\sigma}^0)\right)$ denote the energy gap between the ground and first-excited states under the external field $\bm{h}^{(0)}+\bm{h}$ and letting $M_0= \sum_i \sigma_i$, for any $\bm{\sigma}\in {\Omega^0}^*$, denote the magnetization of the optimal ground states, for sufficiently low temperatures, the magnetization takes the form:
\begin{align}
M(\bm{h}^{(0)}+\bm{h}) &\approx \frac{\left\vert \Omega^0_{\bm{h}}\right\vert M_0 + \left(\sum_{\bm{\sigma}\in \Omega^1_{\bm{h}}} \sum_i \sigma_i\right) e^{-\beta \Delta E(\bm{h})}}{\left\vert \Omega^0_{\bm{h}}\right\vert + \left\vert \Omega^1_{\bm{h}}\right\vert e^{-\beta \Delta E(\bm{h})}} \nonumber \\
&\approx M_0 + c(\bm{h})e^{-\beta \Delta E(\bm{h})},
\end{align}
where $c(\bm{h}) = \frac{1}{\left\vert \Omega^0_{\bm{h}}\right\vert}\left[\left(\sum_{\bm{\sigma}\in \Omega^1_{\bm{h}}} \sum_i \sigma_i\right)- \left\vert \Omega^1_{\bm{h}}\right\vert M_0 \right]$ is a scalar that depends on $\bm{h}$. Thus, the low-temperature optimal external fields, i.e., the external fields which maximize the low-temperature magnetization, take the form:
\begin{align}
\label{LowTh2}
\bm{h}^* &= \argmax_{\bm{h}\in \mathcal{F}_H^*} c(\bm{h}) e^{-\frac{1}{T}\Delta E(\bm{h})} \nonumber \\
&\equiv \argmax_{\bm{h}\in \mathcal{F}_H^*} -\sign\left[c(\bm{h})\right] \Delta E(\bm{h}) \nonumber \\
&\equiv \argmax_{\bm{h}\in \mathcal{F}_H^*}\min_{\bm{\sigma}\in \Omega^1} \max_{\bm{\sigma}^0\in {\Omega^0}^*}  -\sign\left[c(\bm{h})\right] \left(E_{\bm{h}}(\bm{\sigma}) - E_{\bm{h}}(\bm{\sigma}^0)\right)
\end{align}

We first note that Eq. (\ref{LowTh2}) is not linear in $\bm{h}$, and hence the useful algorithmic properties of Eq. (\ref{LowTh}) have been lost in generalizing to general Ising systems. However, despite being non-linear in $\bm{h}$, Eq. (\ref{LowTh2}) does reveal a major insight into the structure of low-$T$ optimal external fields in general Ising systems. Depending on the sign of $c(\bm{h})$, the optimal external field $\bm{h}^*$ will either maximize or minimize the energy gap. Since the first excited states in $\Omega^1_{\bm{h}}$ are likely ground states under other external fields (i.e., it is likely that $\Omega^1_{\bm{h}} \subset \Omega^0$ for $\bm{h}\in\mathcal{F}_H^*$) and since $M_0$ is the largest magnetization among the states in $\Omega^0$, we should expect in most cases that $c(\bm{h}) < 0$ for all $\bm{h}\in \mathcal{F}^*_H$. In this case, the low-$T$ optimal external field $\bm{h}^*$ maximizes the energy gap $\Delta E(\bm{h})$, focusing $H$ on the loosely-connected nodes, i.e., nodes with opposite parity between the ground and first-excited states. Thus, once we restrict our attention to external fields that induce ground states with the maximum magnetization (i.e., once we restrict to $\bm{h}\in\mathcal{F}_H^*$), much of the intuition developed in Section \ref{LowT} generalizes naturally to general Ising systems.

\bibliographystyle{abbrvnat}
\bibliography{IsingInfluenceBib}

\end{document}